\documentstyle[11pt,bezier]{article}
\input{prepictex.tex}
\input{pictex.tex}
\input{postpictex.tex}
\title{Relativistic quark model and scalar diquarks charge radii}
\author{S.M.~Gerasyuta\thanks{Present address: Department of Physics,
 LTA, Institutski Per.5, St.Petersburg 194021, Russia }  ,
 D.V.~Ivanov \\
{\it Department of Theoretical Physics,}\\ {\it St. Petersburg
 State University}\\ {\it 198904, St.Petersburg, Russia }}

\date{}
\begin{document}

\maketitle

\begin{abstract}

In the framework of relativistic quark model the behaviour of
 electromagnetic form factors of diquarks with $ J^{P}=0^{+} $ at small and
 intermediate momentum transfer are determined.
 The charge radii of nonstrange and strange scalar diquarks are calculated.

\end{abstract}

\section{Introduction}
\label{intro}

        Diquarks have become now an efficient tool for studying various
 processes in hadron physics. Diquark model arises naturally if we assume,
 that the strong two-quark correlations determine the properties of baryons.
 As a result, baryons can be considered as the connected quark-diquark states
 [1-3].

        As a first step towards describing baryon form factors, one calculate
 the on-shell electromagnetic form-factors of the constituent diquarks. The
 diquark form factors are important ingredient in the baryon form factor and
 contain information about the sizes of the correlated diquark states.
 Evidence for correlated diquark states in baryons is found in deep-inelastic
 lepton scattering [4] and in hyperon weak decays [5].

        In the framework of the dispersion N/D-method with help of the
 iteration bootstrap procedure the scattering amplitudes of dressed quarks
 were constructed [6]. The mass values of the lowest mesons ($ J^{PC}=
 0^{-+}, 1^{--}, 0^{++} $) and their quark content are obtained.
 The $qq$-amplitudes in the colour state $ {\bar{3}}_{c} $ have the diquark
 levels with $ J^{P}= 0^{+}$ and the masses $ M_{ud}=$0.72~GeV, $M_{us}=M_{ds}=$
 0.86~GeV.

     The present paper is devoted to the calculation  of electric
 form factors and charge radii of nonstrange and strange scalar diquarks.
 The dispersion relation technique allows us to consider the relativistic
 effects in the composite systems. The double dispersion relations over the
 masses of the composite particles are used for the consideration of
 the electric scalar diquark form factors in the infinite momentum frame
 (section 2). In the Conclusion the calculation results for the electric
 form factors, charge radii of scalar diquarks and the status of the
 considered model are discussed.

\section{Electric diquark form factors in the infinite momentum frame}
\label{sect:1}

 We consider electromagnetic form factor of two-quark system.
 Let the masses of quarks composing the system be equal $ m_{1} $
 and $ m_{2} $ respectively. The Feynman amplitude for process of scattering
 of virtual photon on the diquark with $ J^{P} =0^{+} $ is described by the
 standard triangle diagram (Fig.1) on which the photon interacts with
 each of the two quarks. This amplitude is equal:

$$ A(q^{2})=
\int \frac{d^{4} k_{2}} {i{(2\pi )}^{4}}
    \frac{Tr[{\gamma}_{5}(\hat{ k_{1}}+m_{1}){\gamma }
    _{\mu}({\hat{k'_{1}}} +m_{1}){\gamma }_{5}
    ( -\hat{P}+\hat{k_{1}}+m_{2})]}{(m^{2}_{1}-k^{2}_{1})
 (m^{2}_{1}-k'^{2}_{1})(m^{2}_{2}-{(P-k_{1})}^{2})}\times $$
\begin{equation}  G({(k_{1}-k_{2})}^{2})
 G({(k'_{1} -k_{2})}^2) {\varepsilon}_{\mu} e_1 f_1
 (q^2) + [1\longleftrightarrow 2 ],  \end{equation}

\noindent
 where the latter contribution correspons to the diagram with particles
 1 and 2 rearranged among themselves. $G$ is the diquark vertex function.
 Here $ e_{1,2} $ and $ f_{1,2}(q^2) $ are charges and the form factors of
 quarks $ m_1 $ and $ m_2 $ respectively.
 Then we obtain:

 $$ A ( q^{2} )=\frac{1}{2} \int \frac{d^{4} k_{2}} {i{(2\pi )}^{4}}
 \left[(P'^{2}-{(m_{1}-m_{2})}^{2}) k_{1\mu }+ \right. $$
 $$\left. (P^{2}-{(m_{1}-m_{2})}^{2}) k'_{1\mu }+q^{2} k_{2\mu }\right] \times $$
 $${\left[(m^{2}_{1}-k^{2}_{1})(m^{2}_{1}-k'^{2}_{1})
 (m^{2}_{2}-{(P-k_{1})}^{2})\right]}^{-1}\times  $$
\begin{equation}  G({(k_{1}-k_{2})}^{2}) G({(k'_{1}-k_{2})}^{2})
 {\varepsilon}_{\mu} e_1 f_1 (q^2) + [1\longleftrightarrow 2 ]  \end{equation}

 Using the expression [7]:

 $$ k_{1\mu } + k'_{1\mu }=\alpha (P_{\mu }+P'_{\mu })+\beta q_{\mu}+
 {(k_{1\mu }+k'_{1\mu })}_{\bot }, $$
\begin{equation}  \alpha=\frac { P'^{2} + P^{2} + 2 {m}_
 {1}^{2}-2 {m}_{2}^{2}-q^{2}} {2 ( P^{2} + P'^{2})-q^{2}-{(P'^{2}-P^{2})}^{2}
 / q^{2}},  \end{equation}
$$ \beta=-\frac{\alpha (P'^{2}-P^{2})}{q^{2}}, $$
 we calculate the amplitude $ A(q^2) $:

\begin{equation} A(q^{2})={\varepsilon }_{\mu} (P_{\mu } + P'_{\mu })
 e_{D} G_{D}^{E} (q^{2}) , \end{equation}
 where the diquark form factor is obtained:

 $$ G_{D}^{E}(q^{2})=
 \int \frac{ d^{4} k_{2}}{i{(2\pi )}^{4}} \frac{((1-\alpha){q}^{2} +
 \alpha (P^{2} + P'^{2}-2{(m_{2}-m_{1})}^{2}))}{4(m_{1}^{2}-k_{1}^{2})
 (m_{2}^{2}-k_{2}^{2})(m_{1}^{2}-{k'}_{1}^{2})} \times $$
\begin{equation}   G({(k_{1}-k_{2})}^{2}) G({(k_{1}-k'_{1})}^{2})
 \frac{e_{1}}{e_{D}} f_{1} (q^{2}) +
 [1\longleftrightarrow 2 ],  \end{equation}
 $ {\varepsilon}_{\mu} q_{\mu} =0 $, $ e_{D}$ is the diquark charge.

    We pass to the infinite momentum frame and use the
 dispersion integration over the masses of composite particles [7].
 The momentum of the composite particle (diquark) along the $z$-axis is large,
 $ P_{z}\to \infty $. Hereafter we introduce the notation
 $ P=k_{1} + k_{2}, P^{\prime } =P + q $ for the initial and final state
 momenta ($ P^{2}=s, P^{\prime 2}=s^{\prime } $).
 $ s $ and $ s^{\prime } $ are the initial and final energy of the composite
 system. The double discontinuity defines the form factor of the two-quark
 system (diquark):

\begin{equation}
 G_{D}^{E}(q^{2})=\int\limits_{{ ( m_{1} + m_{2} )}^{2}}^
  {{ \Lambda }_{s}} \frac{ds}{ \pi}\int\limits_{{ ( m_{1} + m_{2
  } ) }^{2}}^{{ \Lambda }_{s}} \frac { d {s}^{ \prime }} {
  \pi}\hskip 2 pt \frac{{disc }_{s} { disc }_{ s^{\prime }} G_{D}^{E}
  ( s, s^{ \prime }, q^{2} ) } { ( s-M_{D}^{2} ) ( s^{ \prime } -
  M_{D}^{2} ) },  \end{equation}

 $${ disc }_{s}{ disc }_{s'} G_{D}^{E} (s, s', q^{2} )=  $$
 $$ \frac{1}{4} G G' \left[e_{1} f_{1} (q^{2})D_{1}(s, s', q^{2})
 {\Delta }_{1}(s, s', q^{2}) + \right. $$
\begin{equation}  \left.  e_{2} f_{2} (q^{2}) D_{2} (s, s', q^{2})
 {\Delta}_{2}(s, s', q^{2})\right]/e_{D},   \end{equation}

 Further we calculate the following terms:

 $$ D_{1}(s_{1}, {s'}_{1}, q^{2} )=\frac{1}{4}[(1-{\alpha }_{1}) q^{2} +
 {\alpha }_{1} ({s'}_{1} + s_{1}-2 {(m_{1}-m_{2})}^{2})] $$
$${\alpha }_{1}=\frac{b_{1} + q^{2} {a}_{1} / s_{1}}
 {2(1-q^{2} {a}_{1}^{2} / s_{1})},
 {\Delta }_{1}(s_{1}, {s'}_{1}, q^{2})=\frac { b_{1} {a}_{1} + 1}{b_{1} +
 c_{1} {a}_{1}} $$

\begin{equation} b_{1}=1 + \frac{m_{1\bot }^{2}-m_{ 2\bot }^{2}}{s_{1}},
 c_{1}=b_{1}^{2}-\frac{4{k}_{\bot}^{2} {\cos}^{2}\phi }{s_1}  \end{equation}

 $$ {a}_{1}=\frac{-b_{1} + \sqrt{(b_{1}^{2}-c_{1})
 (1-s_{1} c_{1} / q^{2})}}{c_{1}}, $$

 $$ s_{1}=\frac { m_{ 1\bot }^{2} + x(m_{2\bot }^{2}-m_{1\bot }^{2})}
 {x(1-x)}, {s'}_{1} =s_{1} + q^{2}(1 + 2 {a}_{1}) $$
 $$ m_{i\bot}^{2} = m_{i}^{2} + k_{\bot}^{2} , i=1,2 $$
 and
\begin{equation} D_{2}=D_{1} ( 1\leftrightarrow 2 ), { \Delta }_{2}=
 {\Delta }_{1} ( 1\leftrightarrow 2 )  \end{equation}

 Finally we obtain:

 $$ G_{D}^{E} (q^{2})=\frac{1}{{(4\pi )}^{3}}\int_{0}^{{\Lambda }_{k_{\bot }}}
 d k^{2}_{ \bot}\int_{0}^{ 2\pi } d\phi
 \int_{0}^{1} dx \frac{1}{x(1-x)} \times  $$
 $$ \sum_{i=1,2 } \frac{ G G'}{(s_{i} -M_{D}^{2})({s'}_{i}- M_{D}^{2})}
 \frac{e_{i}}{e_{D}} f_{i}(q^{2}) \times $$
\begin{equation} D_{i}(s_{i}, {s'}_{i}, q^{2}){\Delta }_{i}
 (s_{i}, {s'}_{i}, q^{2})   \end{equation}

\noindent
 The eq.(10) was used in the calculation of the diquark form factors provided
 the normalization $ G_{D}^{E}(0)=1 $.

\section{Conclusion}
\label{sect:2}

 In the present paper in the framework of dispersion integration technique we
 investigate the behaviour of electric diquark form factors with $ J^{P}=
 0^{+}$ at small and intermediate momentum transfer $ Q^{2} \leq
 0.5~{\rm GeV}^{2}$.  The charge radii values of nonstrange and strange scalar
 diquarks are calculated.
  The scalar diquark masses were calculated [6]:
 $ M(ud)=0.72$~GeV, $M(us)=M(ds)=0.86$~GeV. The quark masses are equal:
 $ m_{u}=m_{d}=0.385$~GeV, $m_{s}=0.510$~GeV.
 Analogously [6] we use the dimensionless pair energy cut-off parameter:
 $ \lambda =12.2 $, that allows us to define the momentum cut-offs:
 $ {\Lambda }_{ k_{\bot}}(qq)= 0.3~{\rm GeV}^{2}$, ${\Lambda }_{k_{\bot }}(qs)
 =0.41~{\rm GeV}^{2} $, where $q=u,d$.
 We consider the interaction of constituent quark with electromagnetic field
 and take into account the nonstrange and strange quark form factors:
 $ f_{q}(q^2)= exp({\gamma}_q q^2)$, ${\gamma}_q =0.33~{\rm GeV}^{-2} $ and
 $ f_{s}(q^2)= exp({\gamma}_s q^2)$, ${\gamma}_s =0.2~{\rm GeV}^{-2} $ [6].
 The behaviour of the scalar diquark electric form factors are shown in Fig.2.
 The calculated charge radii are equal:

$ {<{r}_{ud}^{2}>}^{\frac{1}{2}} = 0.55~fm$, ${<{r}_{us}^{2}>}^{\frac{1}{2}}
 = 0.65~fm$,  ${<{r}_{ds}^{2}>}^{\frac{1}{2}} = 0.5~fm  $.

\noindent
 In the present paper electromagnetic properties of diquarks are investigated
 in the framework of the relativistic description.
 These values of scalar
 diquark charge radii are compatible with other results for the diquark
 effective radii [8-11] and experimental data [1].
 In the papers [8,9] assuming soft symmetry breaking in the diquark sector,
 the bosonisation of a quasi-Goldstone ud-diquark is performed. In the chiral
 limit the ud-diquark mass and diquark charge radius are defined by the gluon
 condensate $ M_{ud}=300$~MeV, ${<{r}_{ud}^{2}>}^{\frac{1}{2}} \simeq 0.5~fm $.
 This model allows to explain the relatively low mass of the scalar diquark.

 A approach is based on a local effective quark model, a Nambu-Jona-Lasinio
 model with a colour-octet current-current interaction [10,11]. One
 calculated the electromagnetic form factors of scalar and axial vector
 diquark bound states using the gauge-invariant proper-time regularization.
 In the paper [11] the scalar diquark masses $ M_{ud} $ and scalar diquark
 charge radii $ {<{r}_{ud}^{2}>}^{\frac{1}{2}} $ for different values of the
 effective diquark coupling constants are calculated. The scalar diquark
 charge radius $ {<{r}_{ud}^{2}>}^{\frac{1}{2}} $ is equal $(0.5-0.55)~fm $.

 But the non-relativistic, QCD-based, potential quark model for the proton
 and neutron inevitably predicts a spin-0 diquark structure with a charge
 radius of the $ 0.35~fm $ or smaller [12,13]. Such conflict between model
 and experimental data might possibly as the influence of relativistic
 effects.

\vspace{7mm}
\noindent
{\Large Acknowledgements}
\vspace{7mm}

\noindent
 The authors would like to thank A.A.~Andrianov, V.A.~Franke, Yu.V.~Novozhilov
 for useful discussions.

\newpage
\begin{figure}[t]
\unitlength 0.6mm
\begin{picture}(135.00,76.67)
\thicklines
\put(25.00,30.00){\line(1,0){110.00}}
\put(25.00,33.00){\line(1,0){25.00}}
\put(50.00,33.00){\line(1,1){30.00}}
\put(80.00,63.00){\line(1,-1){30.00}}
\put(110.00,33.00){\line(1,0){25.00}}
\put(48.33,30.00){\rule{1.67\unitlength}{3.00\unitlength}}
\put(110.33,30.00){\rule{1.67\unitlength}{3.00\unitlength}}
\put(110.00,30.00){\vector(-1,0){30.00}}
\put(50.00,33.00){\vector(1,1){16.00}}
\put(80.00,63.00){\vector(1,-1){16.00}}
\bezier{48}(80.00,63.00)(75.67,67.00)(80.33,70.33)
\bezier{36}(80.00,70.33)(83.33,73.00)(80.33,76.67)
\put(83.67,75.67){\makebox(0,0)[lc]{q}}
\put(35.00,27.33){\makebox(0,0)[ct]{$P$}}
\put(125.67,27.33){\makebox(0,0)[ct]{$P'$}}
\put(49.33,27.00){\makebox(0,0)[ct]{${\gamma}_5 $}}
\put(111.67,27.00){\makebox(0,0)[ct]{${\gamma}_5 $}}
\put(76.00,63.00){\makebox(0,0)[rc]{${\gamma}_{\mu}$}}
\put(61.67,48.00){\makebox(0,0)[rc]{$k_1$ }}
\put(99.00,48.00){\makebox(0,0)[lc]{${k'}_1$}}
\put(80.00,26.67){\makebox(0,0)[ct]{$-k_2$}}
\end{picture}
\vspace*{-0.7cm}
\caption{Triangle diagram which defines the form factor of diquark.}
\end{figure}
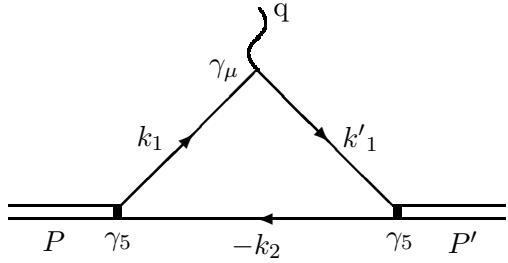


\begin{figure}[b]
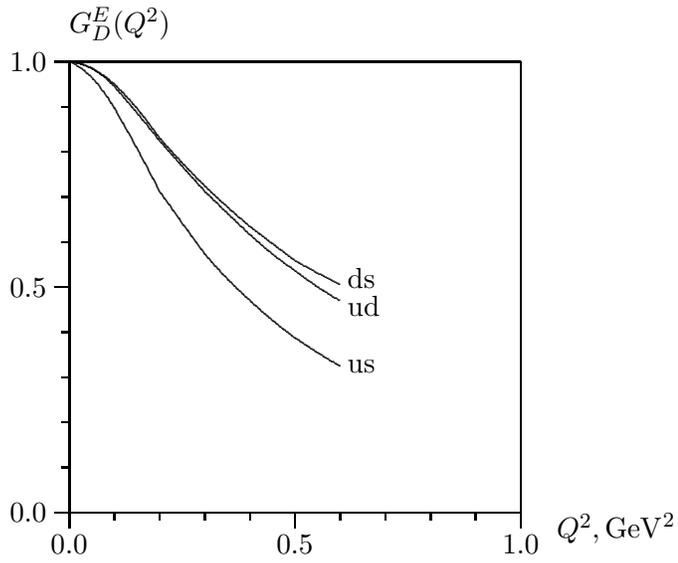

\beginpicture
\setcoordinatesystem units <0.6mm,0.6mm>
\putrule from 0 0 to 0 100
\putrule from 100 0 to 100 100
\putrule from 0 0 to 100 0
\putrule from 0 100 to 100 100

\multiput {\vrule height 0.4pt width 1mm} [r] at 0 0 *10 0 10 /
\multiput {\vrule height 0.4pt width 2mm} [r] at 0 0 *2 0 50 /
\multiput {\vrule height 1mm width 0.4pt} [t] at 0 0 *10 10 0 /
\multiput {\vrule height 2mm width 0.4pt} [t] at 0 0 *2 50 0 /

\put {0.0} [t] <0mm,-3mm> at 0 0
\put {0.5} [t] <0mm,-3mm> at 50 0
\put {1.0} [t] <0mm,-3mm> at 100 0
\put {0.0} [r] <-3mm,0mm> at 0 0
\put {0.5} [r] <-3mm,0mm> at 0 50
\put {1.0} [r] <-3mm,0mm> at 0 100

\put {$G^E_D (Q^2)$} [lt] at 0 112
\put {$Q^2 , {\rm GeV}^2 $} [lt] at 108 0

\put {ds} [lb] <1mm,-0.5mm> at 60 50.6
\put {ud} [lt] <1mm,0.6mm> at 60 47
\put {us} [l] <1mm,0mm> at 60 32.5

\setplotarea x from 0 to 100,
             y from 0 to 100
\setquadratic
\plot 0 100  5 98.5  10 95  15 89.646  20 83.1  25 77.625  30 72.5  35 67.835  40 63.5  45 59.65  50 55.9  55 53.156  60 50.6 /
\setquadratic
\plot 0 100  5 98.5  10 94.5  15 88.666  20 82.5  25 76.854  30 71.3  35 66.443  40 61.7  45 57.394  50 53.7  55 50.098  60 47 /
\setquadratic
\plot 0 100  5 96.5  10 89.7  15 80.809  20 71.3  25 64.288  30 57.4  35 51.898  40 47.1  45 42.675  50 38.8  55 35.505  60 32.5 /

\endpicture
\vspace*{0.5cm}
\caption{The scalar diquarks electric form factor at small and intermediate
momentum transfer $ Q^2 \leq 0.5~{\rm GeV}^2 (Q^2 \equiv -q^2 )$.}
\end{figure}


\begin{thebibliography}{15}

\bibitem{1}
M.~Anselmino  et al.,  Rev. Mod. Phys. {\bf 65},
   1199 (1993)

\bibitem{2}
S.~Fredriksson, M.~J\"{a}ndel ,  T.~Larsson, Z.~Phys. C{\bf 14},
    35 (1982)

\bibitem{3}
H.G.~Dosch, M.~Jamin, B.~Stech, Z.~Phys. C{\bf 42}, 167 (1989)

\bibitem{4}
A.~Donnachie, P.V.~Landshoff, Phys. Lett. B{\bf 95}, 437 (1980)

\bibitem{5}
B.~Stech, Phys. Rev. D{\bf 36}, 975 (1987)

\bibitem{6}
V.V.~Anisovich, S.M.~Gerasyuta, A.V.~Sarantsev, Int. J. Mod. Phys.
 A{\bf 6}, 625 (1991)

\bibitem{7}
V.V.~Anisovich, A.V.~Sarantsev, Sov. J. Part. Nucl. {\bf 45}, 1479 (1987)

\bibitem{8}
Yu.~Novozhilov, A.~Pronko, D.~Vassilevich, Phys. Lett. B{\bf 321}, 425 (1994)

\bibitem{9}
Yu.~Novozhilov, A.~Pronko, D.~Vassilevich, Phys. Lett. B{\bf 343}, 358 (1995)

\bibitem{10}
A.~Buck, R.~Alkofer, H.~Reinhardt, Phys. Lett. B{\bf 286}, 29 (1992)

\bibitem{11}
C.~Weiss et al., Phys. Lett. B{\bf 312}, 6 (1993)

\bibitem{12}
I.M.~Narodetski, Yu.A.~Simonov, V.P.~Yurov, Z.~Phys. C{\bf 55}, 695 (1992)

\bibitem{13}
S.~Fredriksson, J.~Hansson, S.~Ekelin, Z.~Phys. C{\bf 75}, 107 (1997)

\end{thebibliography}
\end{document}